\documentclass[12pt,twoside]{article}
\usepackage{xrb2007}
\def\etal {{\it et~al.}}

\usepackage{natbib,graphics,graphicx,subfigure}

\begin{document}

% select your session by uncommenting the appropriate line
\session{Jets}

\shortauthor{Vrtilek}
\shorttitle{Multiwavelength X-ray binaries}

\title{Multiwavelength Studies of X-ray Binaries}
\author{Saeqa D. Vrtilek}
\affil{Harvard-Smithsonian Center for Astrophysics,
60 Garden Street, Cambridge, MA 02138}

\begin{abstract}
Simultaneous multiwavelength studies of X-ray binaries have
been remarkably
successful and resulted in improved physical constraints,
a new understanding
of the dependence of mass accretion rate on X-ray state,
as well as insights
on the time-dependent relationship between disk structure
and mass-transfer
rate.  I will give some examples of the tremendous gains 
we have obtained in
our understanding of XRBs by using multiwavelength 
observations.  I will end
with an appeal that while Spitzer cryogens are still available 
a special effort be put forth to obtaining coordinated
observations with emphasis on the mid-infrared: Whereas the
optical and near-IR originate as superpositions of the
secondary star and of
accretion processes, the mid-IR crucially detects jet
synchrotron emission
from NSs that is virtually immeasurable at other
wavelengths. A further
benefit of Spitzer observations is that mid-infrared
wavelengths can easily
penetrate regions that are heavily obscured.  Many
X-ray binaries lie in the
Galactic plane and as such are often heavily obscured
in the optical by
interstellar extinction. The infrared component of
the SED, vital to the study
of jets and dust, can be provided {\it only} by Spitzer;
in the X-rays we
currently have an unprecedented six satellites available
and in the optical
and radio dozens of ground-based facilities to complement
the Spitzer
observations.

\end{abstract}

\section{Why Multiwavelength?}

X-ray binaries (XRBs) 
owe their prominence to one of the most efficient energy release
mechanisms known: accretion onto a compact object.
The energy released by accretion can be spread over
essentially the
entire electromagnetic spectrum.  
Since each part of the spectrum provides distinct
and often time-variable information (see, e.g., Fig. 1),
attempts to understand these systems by concentrating on
information from a limited wavelength range may lead
to contradictory and misleading conclusions (see, e.g., Figs. 2-4).
It is important to study them simultaneously over a broad
range of the electromagnetic spectrum;     
spectral energy distributions (SEDS) 
are particularly useful 
because they clearly reveal multiple emission components and 
reflect the physics and geometry of the emitting regions.

Multiwavelength studies have been remarkably successful
and resulted in improved
physical constraints on the systems, a new
understanding of the
dependence of
mass accretion rate on X-ray state,
as well as insights on the time-dependent
relationship between disk structure and mass-transfer rate,
(e.g., Vrtilek~\etal~1990,1991; Hasinger~\etal~1990;
McClintock~\etal~2001; Fuchs~\etal~2003; Homan~\etal~2005)

One of the fundamental unsolved problems of accretion
physics is how the outgoing power is distributed between
electromagnetic radiation and mechanical power (relativistic
jets), and what determines a switch between those two output
channels. 
Relativistic jets are amongst the most energetic phenomena 
in the Universe.
In our Galaxy, jets have been detected in young stellar objects,
massive binaries, symbiotic stars, BH and NS XRBs,
super-soft X-ray sources, and planetary nebulae.
They are also believed to be responsible for $\gamma$-ray bursts
(Fender~\etal~2004).
Despite the tremendous range of scale, jets in stellar mass
systems
share physical properties with those in
super-massive BHS.
The most significant property shared
is that {\it all systems that produce jets also have
accretion disks} (Livio 2002).  This implies that
jet energy is obtained through accretion power and that they
may play a major role in the transport of angular momentum of
the inflalling gas.  As spatial separation is generally
impossible owing to limitation in resolution, {\it multiwavelength 
studies
are essential to separate
the stellar, wind, disk, and jet components.
The broad band spectrum is also necessary to test
theoretical models for jet
formation and to
constrain fundamental parameters of jet physics}.

\begin{figure}
\centering
{
    \scalebox{0.22}{\includegraphics{vrtilek_s_fig1.ps}}
}
{
    \scalebox{0.30}{\includegraphics{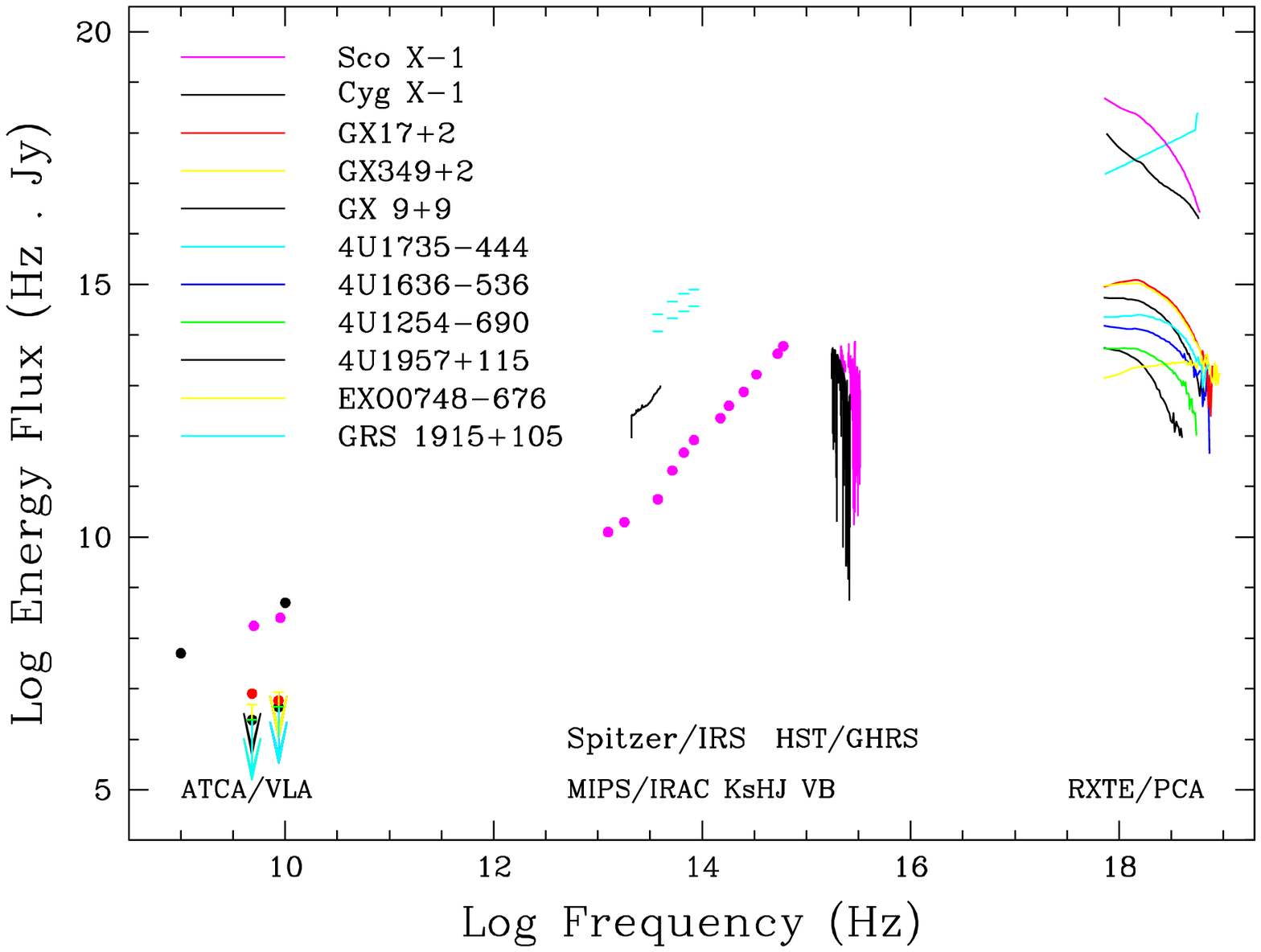}}
}
\caption{
LEFT: Broadband energy spectra of the black-hole binary GRO~J1655-40
illustrating variability within one source.
Each color represents simultaneous observations. (Migliari~\etal~2007;
see also Tomsick contribution to these proceedings).
RIGHT: Broadband energy spectra of a selection of black hole and
neutron star binaries showing variability between sources  
(ATCA and RXTE values are from a survey conducted by the Disk/Jet
Consortium in June of 2006; 
values for Sco X-1 are from Wachter~\etal~2006).
}
\label{fig1sub}
\vspace{-3mm}
\end{figure}

\begin{figure}
\centering
{
    \scalebox{0.3}{\includegraphics{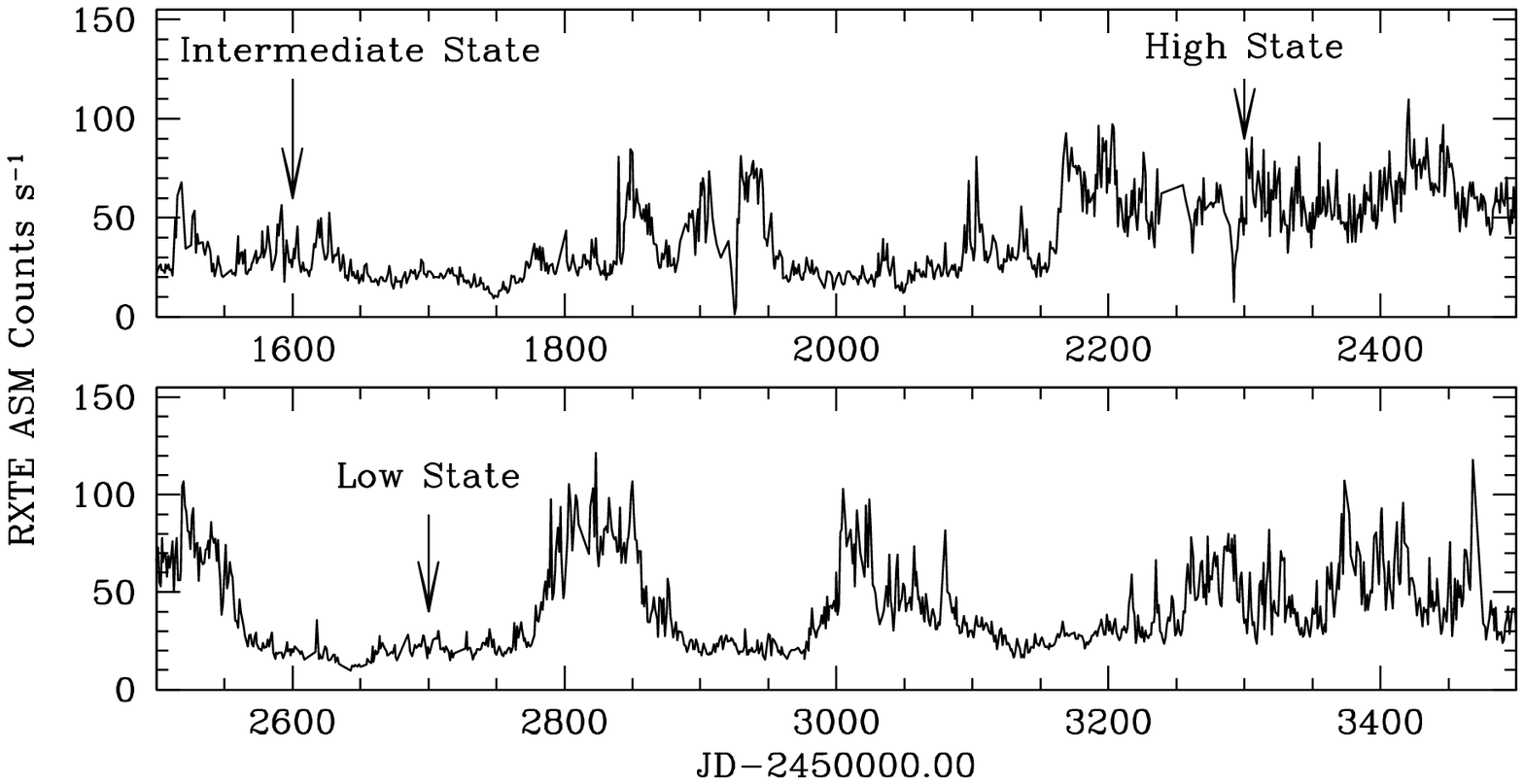}}
}
{
    \scalebox{0.25}{\includegraphics{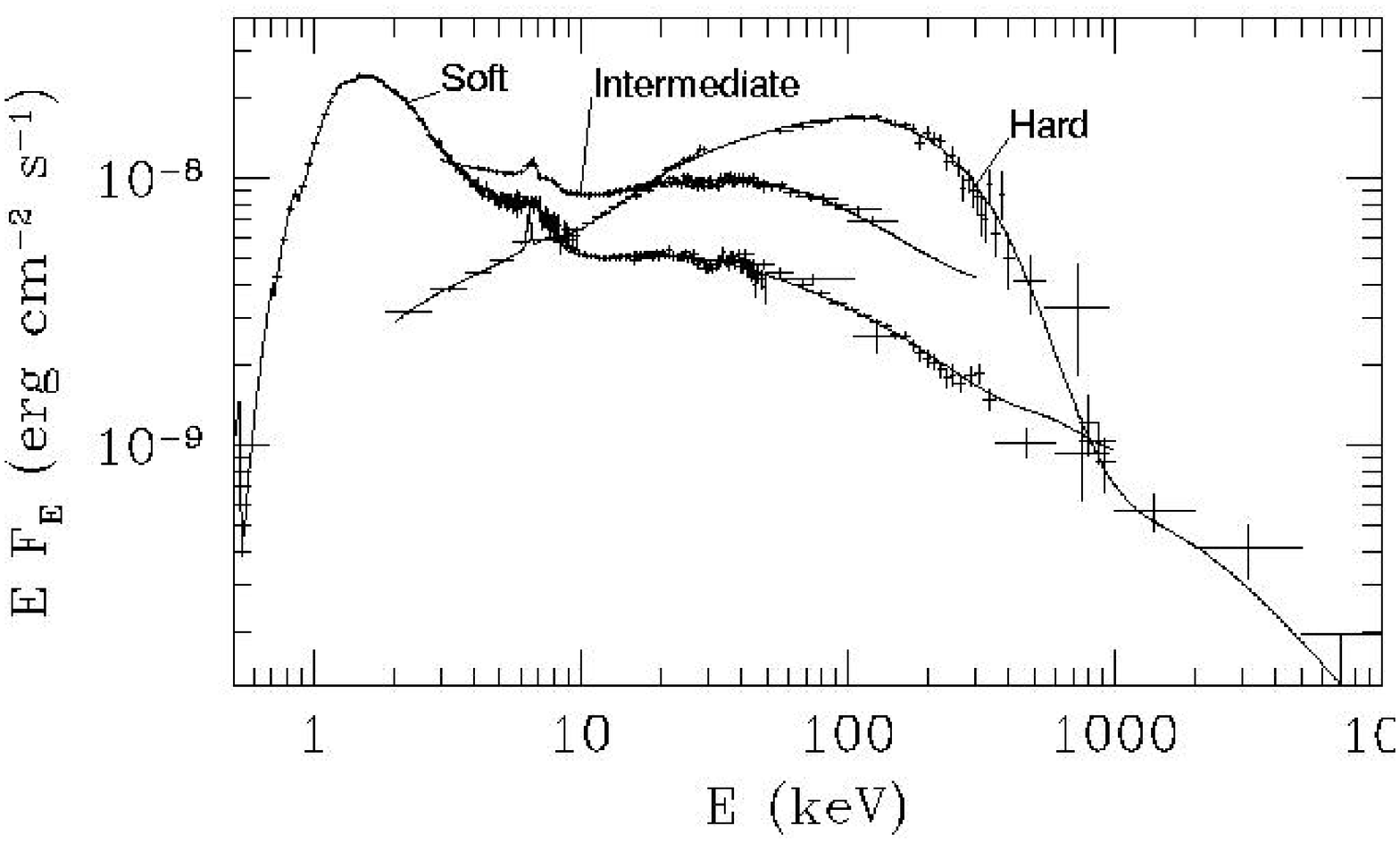}}
}
\caption{
LEFT:
RXTE ASM countrates (courtesy of the RXTE ASM team) 
of the black hole candidate Cygnus X-1 
with typical countrates for ``Low", ``Intermediate", and ``High" 
states indicated. 
RIGHT: 
Broadband energy spectrum of Cygnus X-1 (Gierlin'ski~\etal~1993).
It is clear that the designations ``High" and ``Low" states
are based on a narrow energy range (2-10 keV) and are not valid 
when the full X-ray band is available. 
}
\label{fig2sub}
\vspace{-3mm}
\end{figure}

\begin{figure}
\centering
{
    \scalebox{0.23}{\includegraphics{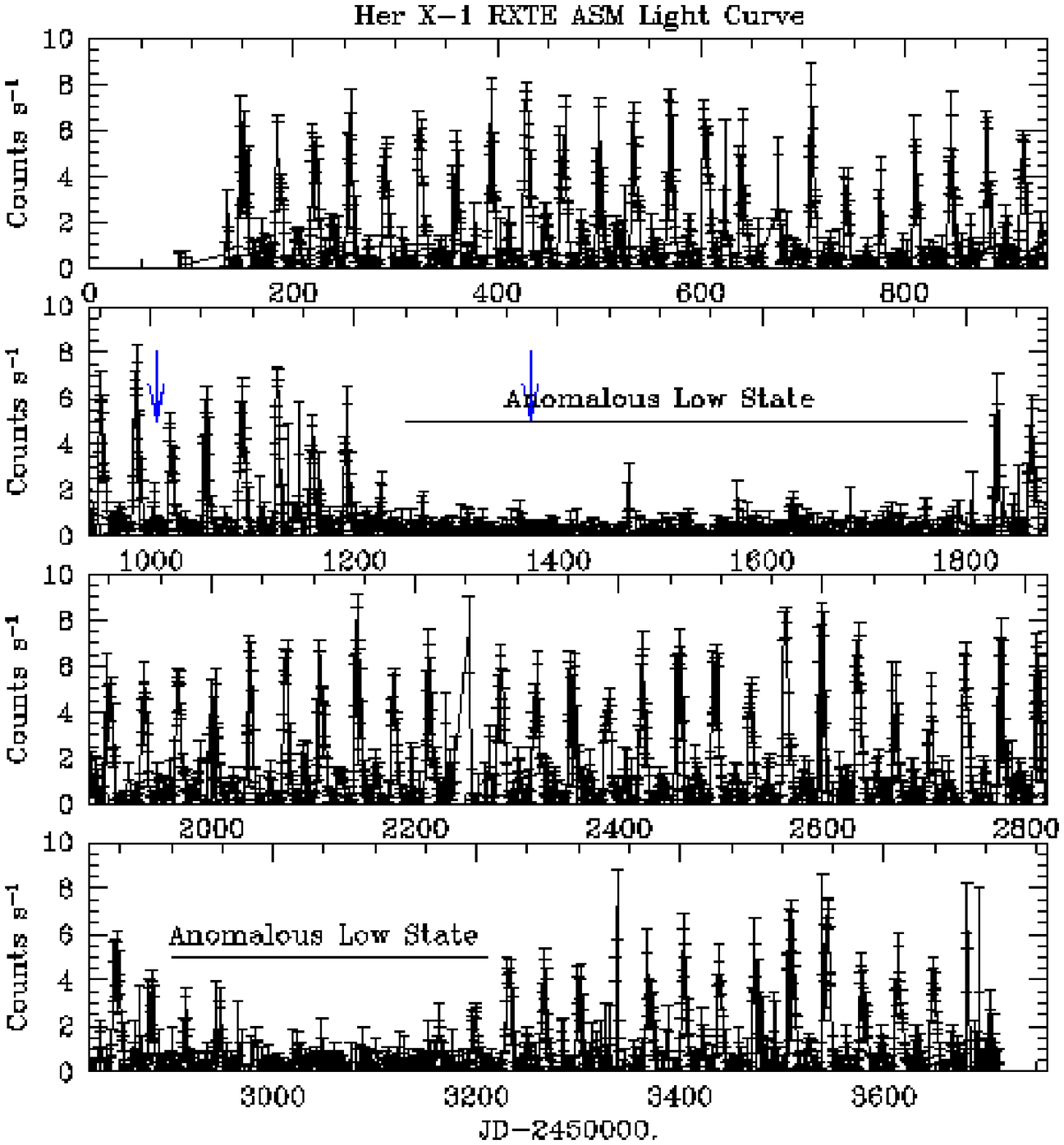}}
}
{
    \scalebox{0.32}{\includegraphics{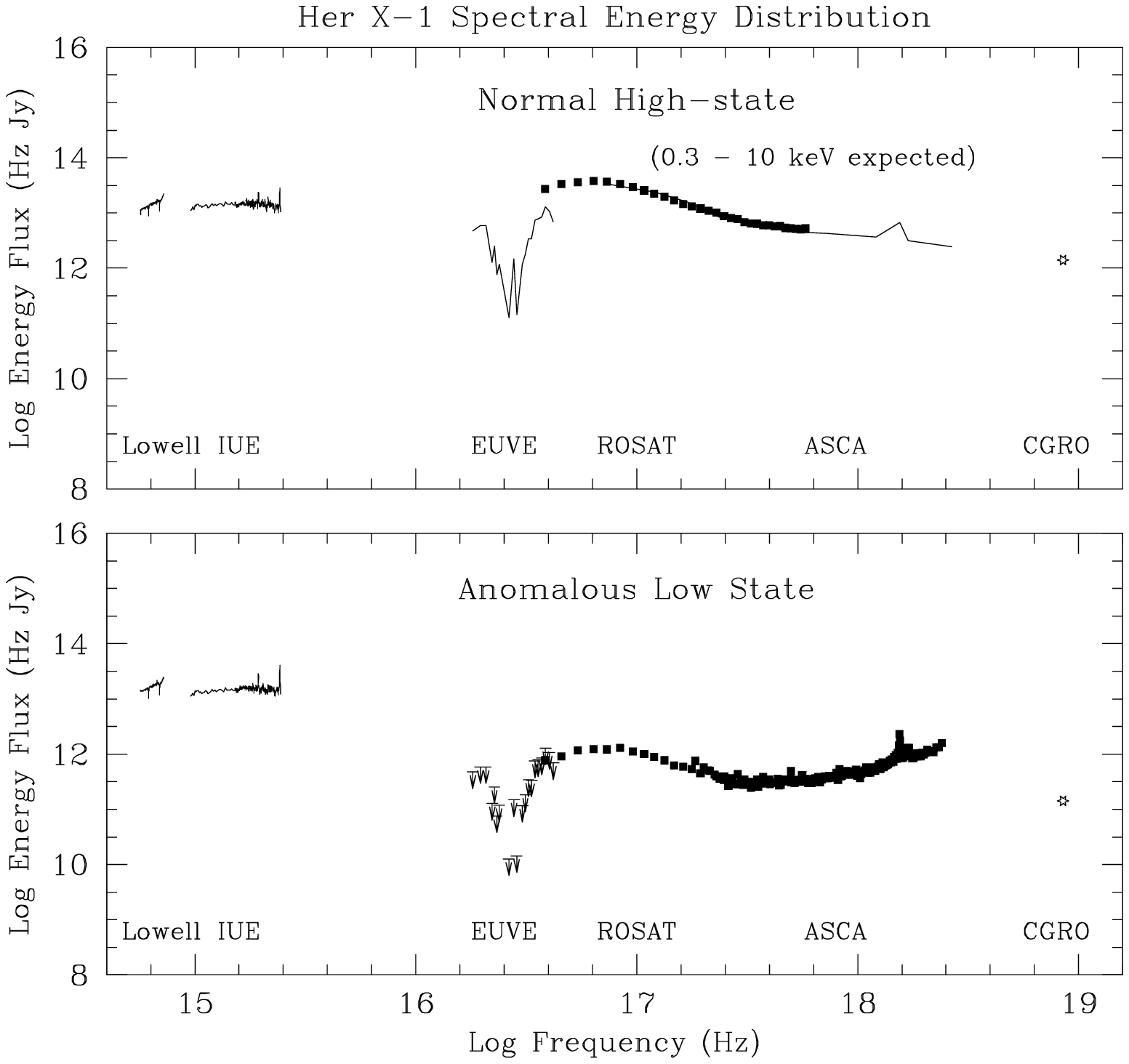}}
}
\caption{
LEFT:
RXTE ASM countrates (courtesy of the RXTE ASM team)
of the pulsing XRB Her X-1.
Arrows mark the dates of the SEDs plotted on the right.
RIGHT:
Simultaneous SEDs of 
Her X-1
clearly show that the optical and ultraviolet flux
(left side) remain largely
unaffected at times when the extreme ultraviolet to hard X-ray
flux (right side) takes a dramatic plunge.  An indication
that X-ray flux is not a dependable measure of mass accretion
rate.
Vrtilek~\etal~2001).
}
\label{fig2sub}
\vspace{-3mm}
\end{figure}

\begin{figure}
\centering
{
 \scalebox{0.18}{\includegraphics{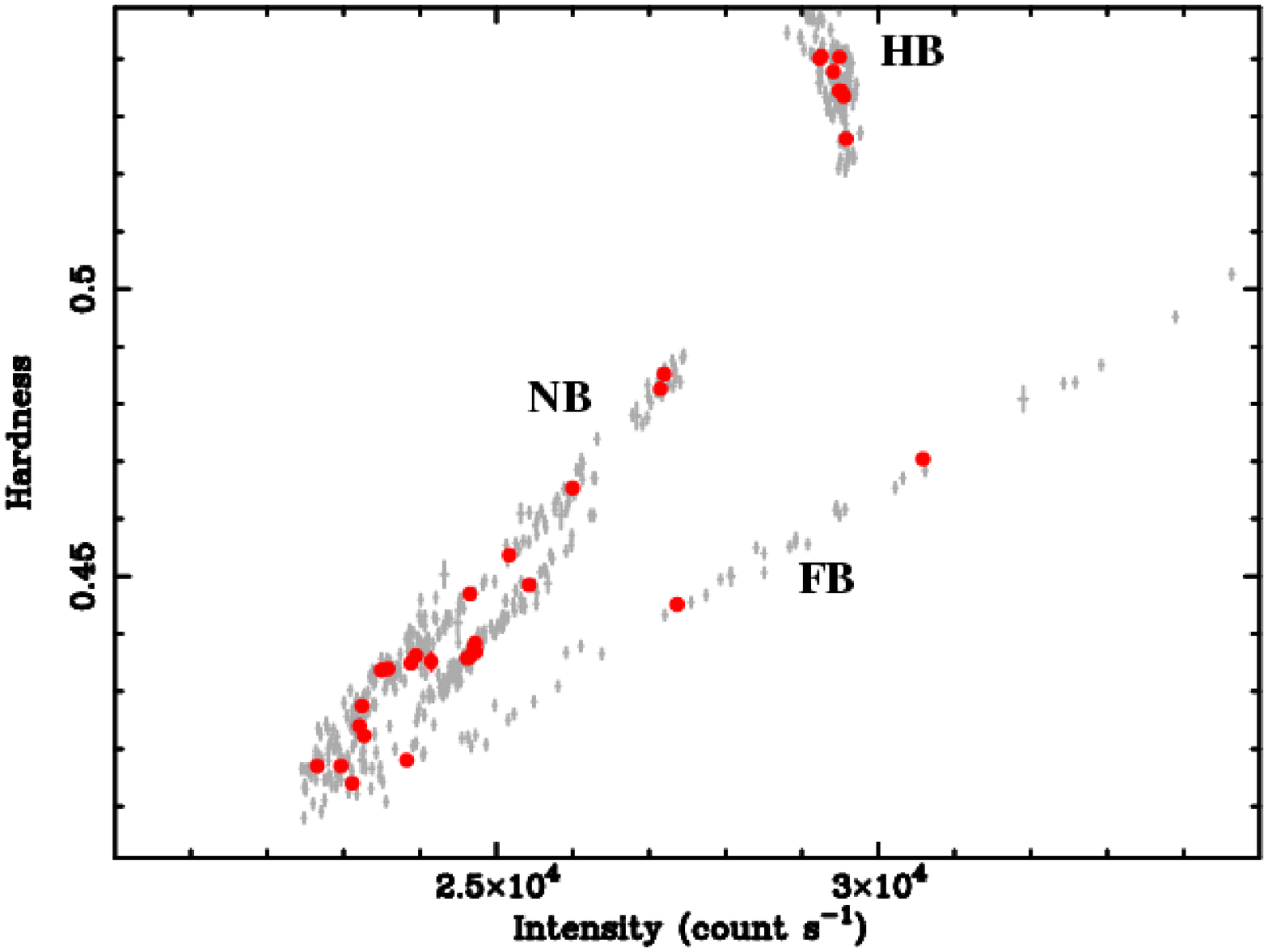}}
}
\hspace{1cm}
{
    \scalebox{0.35}{\includegraphics{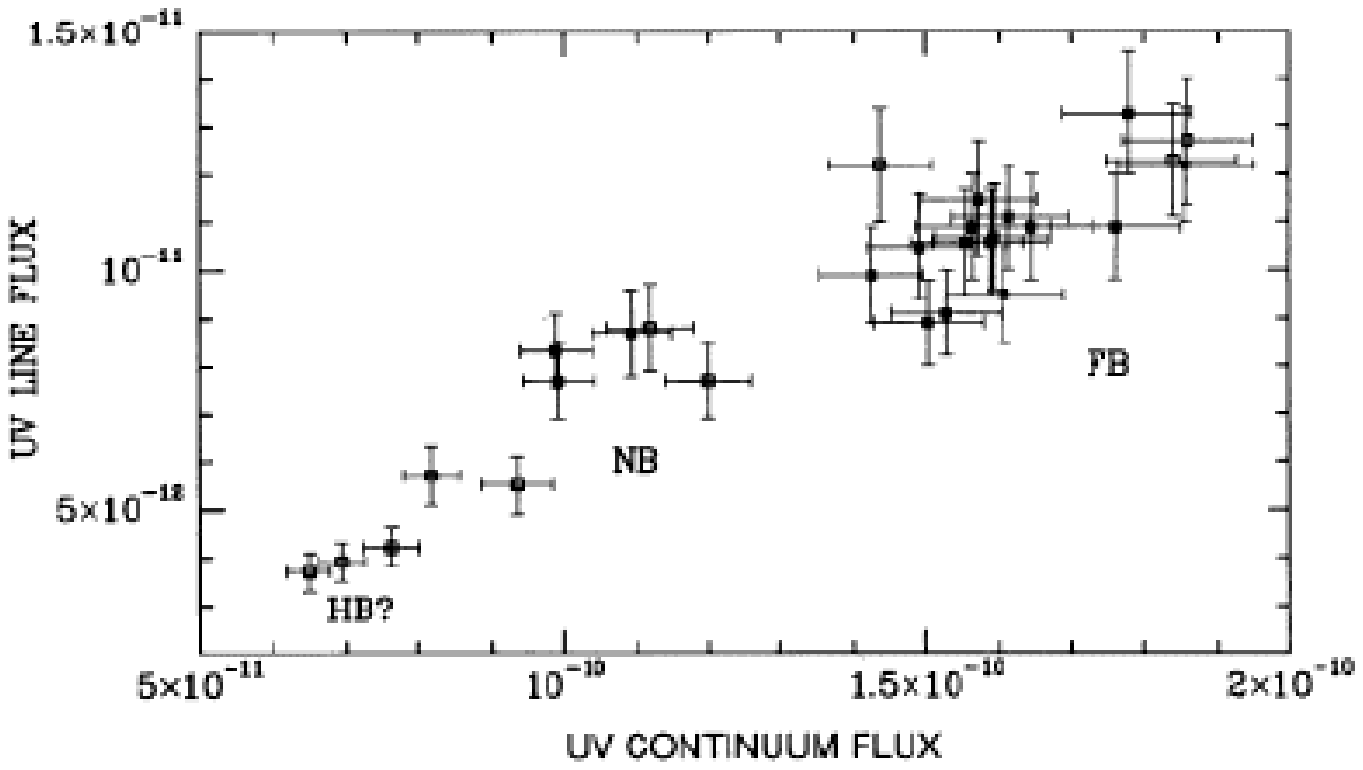}}
}

\caption{
LEFT: 
Hardness vs. intensity data from RXTE observations of the Galactic 
Z-source Sco X-1  
(Barnard, Kolb, \& Osborne 2003).
RIGHT: UV continuum vs line flux in 1224-1986$\AA$ band of
Sco X-1
verifying that mass accretion rate increases from HB to
NB to FB and does not correlate with X-ray flux. (Vrtilek~\etal~1991).} 
\label{cygscocyg}
\end{figure}

\section{Why Spitzer?}

\begin{figure}
\centering
{
    \scalebox{0.3}{\includegraphics{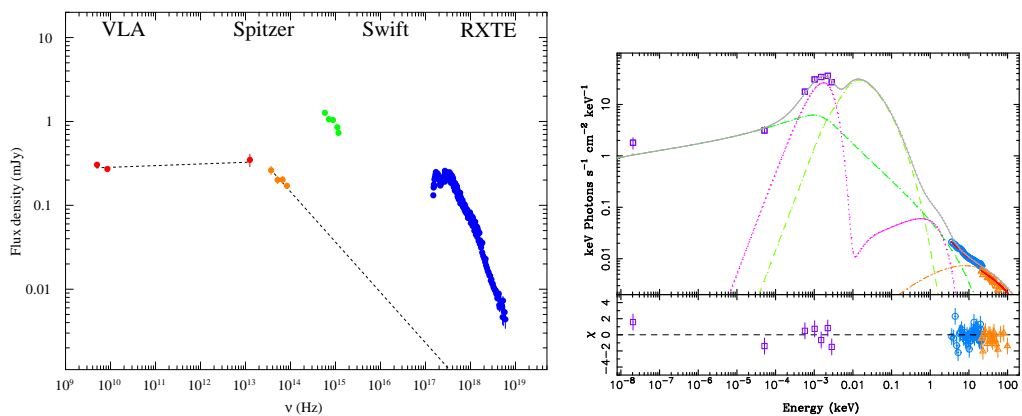}}
}
{
    \scalebox{0.3}{\includegraphics{vrtilek_s_fig10.ps}}
}
\caption{
Left Panel:
Spectrum of the NS 4U 0614+091
showing the clear break defined by IRAC and MIPS observations
(Migliari~\etal~2006,2007).
Right panel:
Jet model fits with residuals
for observation of GRO~J1655-40, with
the jet inclination angle fixed to $75^{\circ}$. The light green
dashed line is the pre-shock synchrotron component,
the darker green dash-dotted
line is the post-shock synchrotron component, the orange dash-dotted line
represents the SSC plus the disk external Compton component,
the purple dotted line is the multi-temperature disk black body plus a
black body representing the companion star in the binary system. The
solid red line is the total model.
(Migliari~\etal~2007).}
\label{fig3sub}
\vspace{-3mm}
\end{figure}

Spitzer observations of XRBs show an amazing variety of behavior:
$\bullet$ The first spectral evidence for a compact jet
in a low-luminosity NS (Migliari~\etal~2006; Fig. 5a).
$\bullet$ The
first detection
of 24$\mu$m  emission from the compact jet of a BH X-ray
binary in its hard state (Migliari~\etal~2007; Fig. 5b).
$\bullet$ Excess mid-IR emission from BH systems
interpreted as
cool dust from circumbinary disks (Muno \&
Mauerhan 2006).
$\bullet$
Multiple continuum, absorption, and emission features
from IGR J16318-4848 which suggest
that the highly-obscured massive systems represent a previously
unknown evolutionary phase of X-ray binaries
(Moon~\etal~2007).
$\bullet$ A large conical cavity surrounding
LMC X-1
which can be attributed to a
shock driven by an as yet unobserved jet from LMC X-1
(Cooke~\etal~2007).
$\bullet$
Emission from dust heated by the shocks in the HMXB Vela X-1
providing a direct measure
of the energy input into the ISM as well as information
on the composition of the dust
(Iping, Sonneborn,
\& Kaper 2007).
$\bullet$ The first detection of a disk around an anomalous X-ray
pulsar (AXP; Ertan~\etal~2007).

Whereas the optical and near-IR
originate as superpositions of the secondary star
and of accretion processes, the mid-IR
crucially detects jet synchrotron emission from NSs
that is virtually unmeasurable at other wavelengths
NS systems have over an order
of magnitude lower radio luminosity (compared to BH systems) making
the IR crucial for detection of jets from these systems.
A further benefit of Spitzer observations is that
mid-infrared wavelengths can easily
penetrate regions that
are heavily obscured.
Many X-ray binaries lie in the Galactic plane
where the interstellar
extinction in the optical is very high.
Spitzer has opened a promising new channel
for the study of
accreting binaries: 
The infrared component of the SED is vital to the study of jets, dust,
 and
obscured systems; {\it only}
Spitzer provides the combination of sensitivity and wavelength
coverage to meet the proposed science goals.
Spitzer IRAC goes 10 times deeper than the next proposed IR mission (WISE) 
and is our only chance
for mid-IR detections of several types of XRBs.

\section{A Website for X-ray Binary SEDs.}

SEDs of stars and AGN
have been used to great
benefit furthering our understanding
of their energy balance, characteristics, and evolution.
Hitherto the SEDs of X-ray binaries have
not been systematically analyzed because the
extreme variability of XRBs
requires
simultaneous observations across the spectrum, and this has been
accomplished for only
a handful of objects (e.g., Fig. 3b).
Cooperative allocation of observing time by many observatories
has increased the ease with which two or even three wavelength studies
can be conducted but getting coordinated coverage over the full available
spectrum continues to be very difficult.

Fortunately, in X-rays the variability can be resolved into a finite
set of discrete spectral states, and the ASM 
on RXTE---operating continuously since 1996---is capable of determining
the spectral states of many XRBs.
We can use this information on X-ray states
in combination with the rich archival databases
available
in radio (NRAO archives), infrared (IRSA), optical and ultraviolet
(e.g., ESO, AAVSO, MAST),
and X-ray (e.g., HEASARC),
to construct SEDs as a function of X-ray spectral state.

The SEDs can be used to conduct
studies to both compare object from different classes and to
follow individual objects
across different states:
goals include determination
of the geometry of the emitting region, disk structure, 
density, ionization,
and
departure from
sphericity of the winds from
massive systems, modeling of long-term variations,
and clarification of the disk-jet connection in XRBs.

We are working on a project (funded by a NASA ADP grant and 
Smithsonian Endowment funds) to produce an atlas
of SEDs as a function of X-ray state for each subtype of 
X-ray binary. 
This will be made available for use by the astronomical
community, providing a
resource that should help
catalyze detailed studies of individual systems as well as comparative
studies
between classes of systems
($http://hea-www.harvard.edu/~saku/SED.html$).

We ask the community to help us by sending their SEDs!

\section{References}

\noindent
Barnard, R., Kolb, U., \& Osborne, J.P. 2003, A\&A, 411, 553.

\noindent
Cooke, R., Kuncic, Z., Sharp, R., \& Bland-Hawthorn, J. 2007,
ApJ, 667, 153.

\noindent
Ertan, U., Erkut, M.H., Eksi, K.Y., \& Alpar, M.A. 2007, ApJ,
657, 441.

\noindent
Fender, R.P., 2001, MNRAS, 322, 31.

\noindent
Fender, R.P., Belloni, T.M., \& Gallo, E. 2004, MNRAS, 355, 1017.

\noindent
Fender, R.P.,~\etal~1999, ApJ, 519, L165.

\noindent
Fender, R.P., Gallo, E., \& Jonker, P.G. 2003, MNRAS, 343, 99L.

\noindent
Fuchs, Y.~\etal~2003, A\&A, 409, L35.

\noindent
Gallo, E., \& Fender, R.P., 2005, Memorie della Societa Astronomica
Italiana, 76, 600.

\noindent
Gallo, E., Migliari, S., Markoff, S., Tomsick, J.A., Bailyn, C.D.,
Berta, S., Fender, R.P., \& Miller-Jones. J.C.A. 2007, astro-ph
0707.0028.

\noindent
Hasinger, G. 1990, A\&A, 235, 131.

\noindent
Homan, J., Buxton, M., Markoff, S., Bailyn, C., Nespoli, E.,
\& Belloni, T. 2005, ApJ, 624.

\noindent
Iping, R., Sonneborn, G., \& Kaper, L. 2007, AAS, 210, 1910.

\noindent
Livio, M. 2002, Nature, 417, 125.

\noindent
McClintock, J.~\etal, 2001, ApJ, 555, 477.

\noindent
Gierlin'ski, M., Zdziarski, A. A., Poutanen, J., Coppi, P. S., Ebisawa, 
K., \& Johnson, W. N. 1999, MNRAS, 309, 496.

\noindent
Migliari, S., \& Fender, R.P., 2006, MNRAS, 366, 79.

\noindent
Migliari, S., Tomsick, J.A., Maccarone, T.J., Gallo, E.,
Fender, R.P., Nelemans, G., \& Russell, D.M. 2006, ApJ, 643L, 41.

\noindent
Migliari, S., Tomsick, J.A., Markoff, S., Kalemci, E., Bailyn, C.D.,
Buxton, M., Corbel, S., Fender, R.P., \& Kaaret, P. 2007,
astro-ph 0707.4500.

\noindent
Moon, D., Kaplan, D.L., Reach, W.T., Harrison, F.A., Lee, J., \&
Martin, P.E. 2007, astro-ph 0710.3351.

\noindent
Muno, M.P., \& Mauerhan, J. 2006, ApJ, 648L, 135.

\noindent
Vrtilek,S.D.~\etal~1990, A\&A, 235, 162.

\noindent
Vrtilek, S.D.~\etal~1991, ApJ, 376, 278.

\noindent
Vrtilek, S.D., Quaintrell, H., Boroson, B., Still, M., Fiedler, H.,
O'Brien, K., \& R. McCray, 2001, ApJ, 549, 522.

\noindent
Wachter, S. 2007, AAS, 211, 0305.

\acknowledgements 

I would like to thank the 
Disks/Jets consortium: 

http://www.cfa.harvard.edu/twiki/bin/view/GreatObs/JetsDisk

for their encouragment, support, and active work on this project.
\end{document}